\documentclass[11pt]{article}
\usepackage{moriond,epsfig}
\def\be{\begin{equation}}
\def\ee{\end{equation}}
\def\bea{\begin{eqnarray}}
\def\eea{\end{eqnarray}}

\begin{document}
\vspace*{3.5cm}
\title{Fermions and Bosons in Superconducting Amorphous Wires}

\author{\underline{YUVAL OREG}$^1$ and EUGENE DEMLER$^2$}
\address{$^1$ Department of condensed matter physics, Weizmann Institute of Science, Rehovot 76100\\
         $^2$ Lyman Laboratory of Physics, Harvard University, Cambridge MA 02138 USA}

\maketitle\abstracts{We discuss the destruction of
superconductivity in quasi-one-dimensional systems due to the
interplay between disorder and Coulomb repulsion. We argue that to
understand the behavior of the system one has to study both
fermionic and bosonic mechanisms of suppression of
superconductivity. The  former describes reduction in the mean
field critical temperature $T_c$, while the latter refers to
thermal and quantum fluctuations in the order parameter. A change
in parameters such as wire width and disorder strength
significantly affects both mechanisms.}

\section{Introduction}

The combination of disorder and Coulomb repulsion suppresses
superconductivity in low dimensional samples, and may destroy it
completely. Advanced technics for fabrication of superconducting
wires allow a detailed experimental
study~\cite{SCW:Bezryadin00,SCW:Lau01,SCW:Kociak01} of this
destruction.

Does $R_N$, the normal-state-resistance of the wire at high
temperature, determine the destruction of superconductivity on it
own? Or do parameters such as the amount of disorder, the width of
the wire, its length, the introduction of a magnetic field, and
the environment have additional effects, beside their direct
effect on $R_N$? Can they lead to a transition in the ground state
properties of the system at zero temperature, and to a qualitative
change in the whole structure of the wire resistance curve as a
function of the temperature $T$?

We argue below, that to understand the answers to these questions
we have to follow both the physics of the electrons above the mean
field $T_c$, and of the Cooper pairs (bosons) below it.
\vspace{-0.4cm}
\begin{figure}[h]
\hskip -1cm
 \psfig{figure=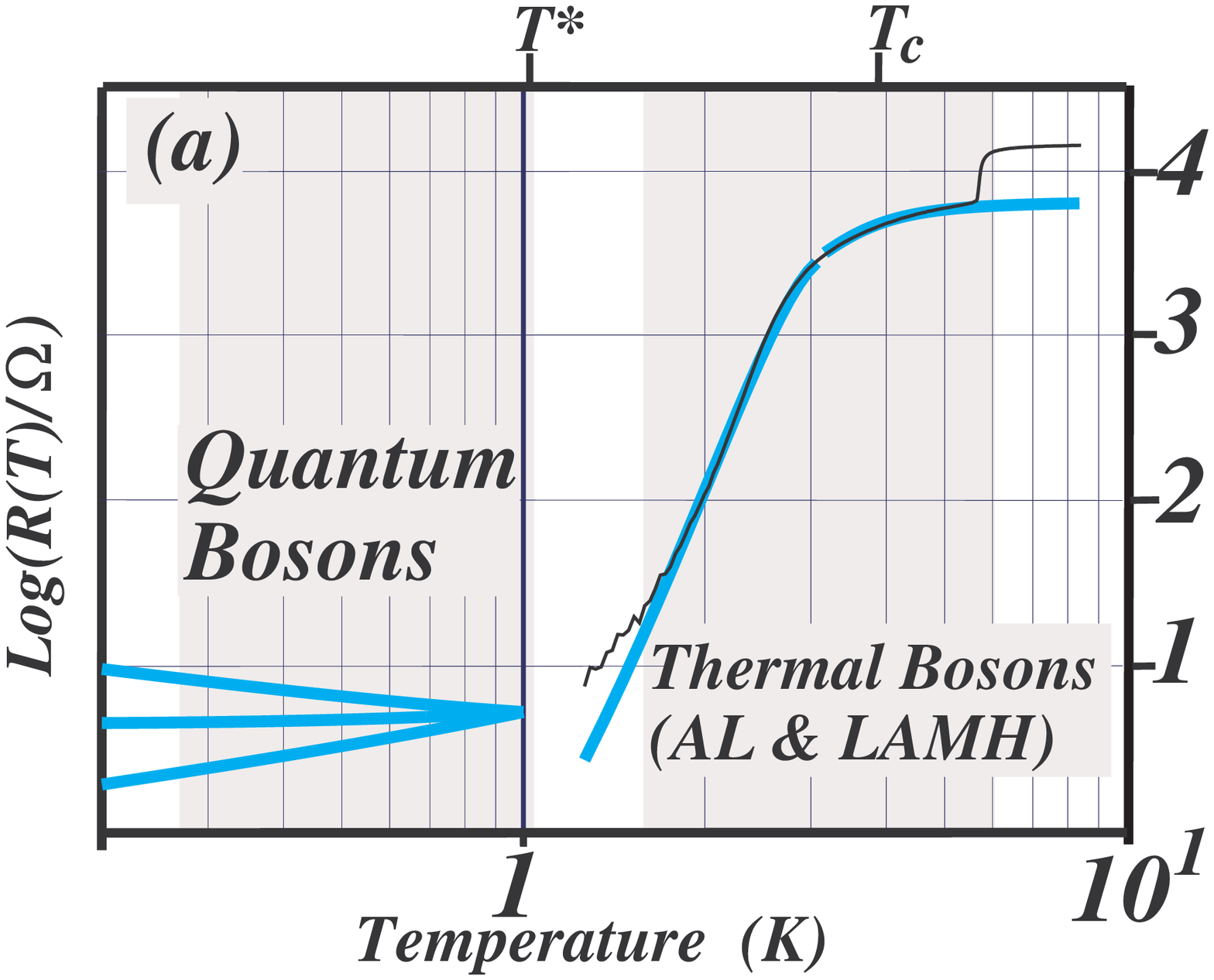,height=3.6in}\
\vskip -3.66in \hskip 3.in \psfig{figure=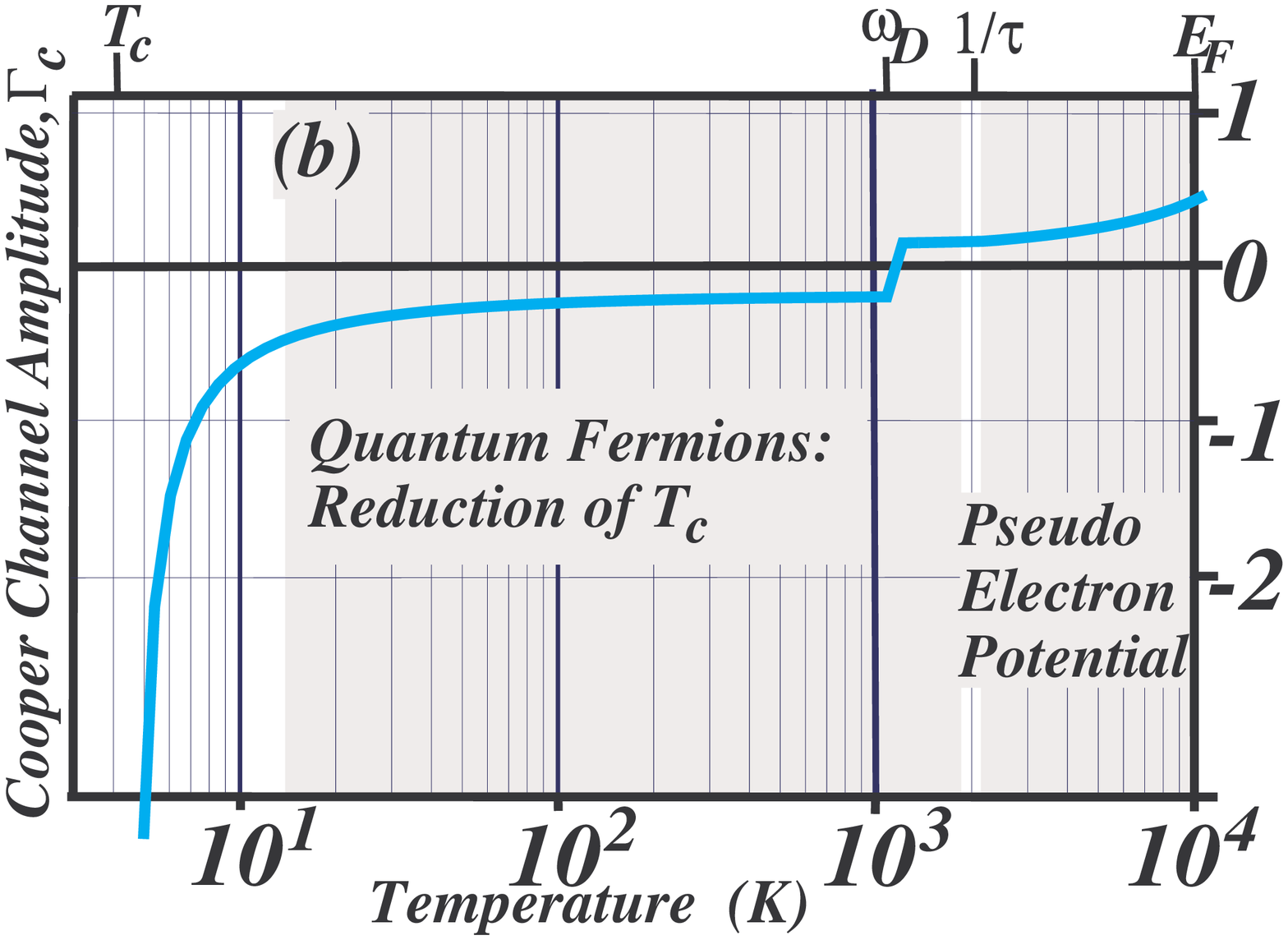,height=3.64in}
 \vskip -3.15cm
 \caption{Theories of fermions and bosons, abbr. are explained in the text,
  notice the $\log$ scale of the temperature. \label{fg:fb}}
\end{figure}
\newpage
   To follow the interplay between bosons and fermions we plot schematically the
evolution of the resistance $R$ (Fig.~\ref{fg:fb}a), and of the
vertex part of the Cooper channel amplitude, $\Gamma_c$
(Fig.~\ref{fg:fb}b), as $T$ decreases. At $T_c$ the
superconducting instability occurs and $\Gamma_c$
diverges.~\cite{RFS:AGD63} It is useful to imagine a flow in the
renormalization-group sense, and find the influence of disorder on
it.

   Let us first discuss $\Gamma_c$. Between the fermi energy $E_F$ and the inverse of the mean free
time between elastic collisions $\tau$, Fermi-liquid parameters
are established. In this temperature range the amplitude
$\Gamma_c$ describes the mutual Coulomb repulsion between
electrons in the Cooper channel. In this region it is reduced by a
logarithmic factor $\propto \log(E_F \tau)$, known as the
Anderson-Morel-Tolmachev $\log$, or {\em pseudo-electron
potential} $\log$.~\cite{SC:Morel62} \hskip 2cm Below $1/\tau$ the
electrons' motion becomes diffusive, "dresses" the mutual Coulomb
repulsion between them, and enhances it. Hence, the
pseudo-electron logarithmic reduction is less
effective.~\cite{SC2D:Finkelstein94,SCW:Oreg99,SCW:Smith01} At the
Debye frequency~\footnote{The case $\omega_D
> 1/\tau$ can be discussed as well but to make the discussion
simpler we assume $1/\tau>\omega_D$.} $\omega_D<1/\tau$ we also
have to include the attraction due to phonons. At yet lower $T$,
$\Gamma_c$ decreases until it diverges at $T_c$.
 Below $\omega_D$ the dynamically dressed Coulomb repulsion changes the whole
evolution of $\Gamma_c$ and reduces $T_c$. It appears that the
dynamic change of the interaction in the Cooper channel arises
from frequencies larger than $T$, we therefore use the phrase {\em
quantum fermions} in Fig.~\ref{fg:fb}b.~\footnote{Below $1/\tau$
there are also corrections to the system properties due to weak
localization and interaction effects. For weak disorder, however,
they are small compared to the correction to $\Gamma_c$ and
therefore we neglect them.}

 Now we move on to discuss $R(T)$ (Fig~\ref{fg:fb}a). Close to, but above $T_c$ we can still describe the system by
fermions and discuss  Cooper pair fluctuations that enhance the
conductance (the Aslamazov-Larkin (AL)
correction).~\cite{DR:Aslamazov68} Alternatively we can describe
the system by fluctuations of a bosonic field (the
complex-order-parameter) with zero average.~\cite{SCW:Tucker71}
Below $T_c$ the magnitude of the order parameter develops and it
is more appropriate to describe the system in terms of an
effective bosonic theory -- the time dependant Ginzburg-Landau
theory. Langer, Ambegaakor, McCumber and Halperin (LAMH)
\cite{SCW:LAMH6770} discuss the way thermal activation of phase
slips in the order parameter influence the resistance of the
system below $T_c$.

At $T \ll T_c $ the amplitude of the order parameter is well
developed and thermal activations of phase slips are very rare.
One then has to consider, the quantized nature of the phase slips,
at frequencies larger than $T$.~\cite{SCW:Tinkham96a} A variety of
phase-only-models were suggested to describe the behavior of the
system at low temperatures. We emphasize that the parameters in
these models depend on fermionic physics which in turn depends on
external parameters.

The resistance increases with the number of phase slips. Coupling
to a dissipative environment increases the action of a single
phase slip and can reduce their number, see for example
Ref.~\cite{SCW:Zaikin97,SCW:Demler01}. The three curves in
Fig.~\ref{fg:fb}a at low $T$ indicate the possible ways the
resistance curve can behave: for strong coupling to a dissipative
environment the number of phase slips (and hence resistivity)
decreases when $T$ decreases, and vice versa for weak coupling to
the environment. The middle curve schematically describes a
possible intermediate case.

For the most part the literature discusses the bosonic or the
fermionic mechanisms separately. Larkin \cite{SC:Larkin99}
addresses the necessity to discuss both of them together in the
case of two dimensional systems. Here we suggest a scheme that
combines fermionic and bosonic physics in quasi-1D.

\section{Comparison with experiment}

 The reduction of $T_c$ due to the enhanced Coulomb repulsion in dirty
wires may be written, in terms of parameters that are
experimentally accessible,
as~\cite{SC2D:Finkelstein94,SCW:Oreg99,SCW:Smith01}
\begin{equation}
\label{eq:Tcsup}
 T_c=T_c^0 \exp\left[-b (4 R_q)/{R_\xi} +
O\left(\left(R_q/R_\xi\right)^2\right)\right],
\end{equation}
where $T_c^0$ is the temperature where $\Gamma_c$ diverges in the
absence of disorder, $R_q =2 \pi \hbar/(2 e)^2 \approx 6.453
k\Omega$ is the quantum resistance, $R_\xi=\rho \xi/A$ is the
resistance for a wire of coherence length $\xi$, and $A$ is the
wire cross section. In the 1D limit where $\xi \gg \sqrt{A}$ the
number $b = \sqrt{A} \Sigma_2 /(\xi \pi)$ is $O(1)$ and depends
weakly on $T_c^0$ and on $A$. In the intermediate region, where
$\xi \sim \sqrt{A}$, $b$ is very sensitive to microscopic
parameters and boundary conditions.~\cite{SCW:Smith01} Notice that
the factor $4R_q$ in Eq.~(\ref{eq:Tcsup}) suggests that we are
dealing with the effects of fermions. The sum (over certain
diffusion propagators) $\Sigma_2$, is defined precisely in
Ref.~\cite{SCW:Oreg99}. To get a good estimate for
$T_c$-reduction, we need to find $b$ accurately since $T_c$
depends on it exponentially, and to include effects of higher
orders of $R_q/R_\xi$. This was performed for a few
examples,~\cite{SCW:Oreg99,SCW:Smith01} by a solution of an
integral equation, .

\begin{figure}[h]
\hskip 2cm \psfig{figure=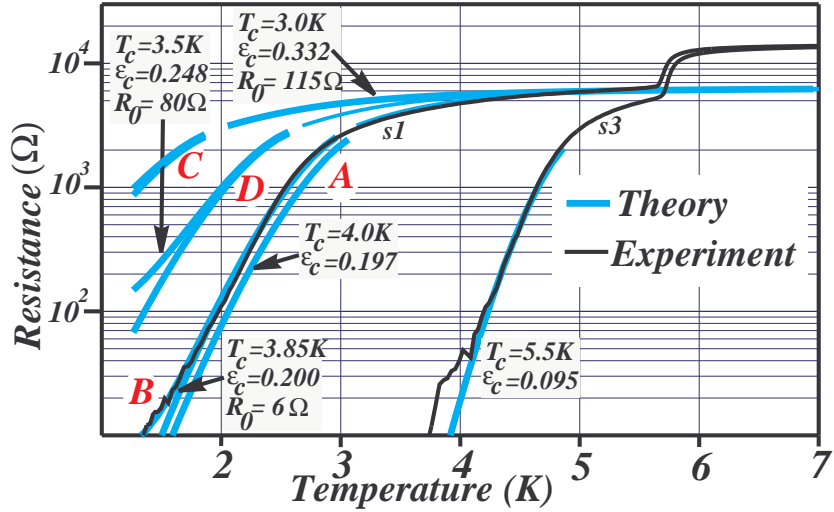,height=3in} \vskip -1cm
 \caption{Comparison between theory and experiments.~$^{16}$\label{fg:et}}
\end{figure}

In Fig. \ref{fg:et} we plot experimental results taken from
Ref.~\cite{SCW:Bezryadin00} in comparison with our theory. We use
the expressions in Ref.~\cite{SCW:Tucker71} to plot the
theoretical resistance curves (gray) based on LAMH-theory
\cite{SCW:LAMH6770} ($T<T_c$),  with the AL correction ($T>T_c$).
LAMH-theory gives:
\begin{equation}
R ( T) =R_N  \; 2^{1/4} \sqrt{\pi} \alpha_0 \left|
\epsilon/\epsilon_c \right|^{9/4} \exp\left(-(4 \sqrt{2}/3) \left|
\epsilon/\epsilon_c \right|^{3/2} \right),
\end{equation}
 with $ \alpha_0 \propto T_c \tau, \;\; \epsilon_c^{3/2} \propto 1/A, \epsilon \propto
 T-T_c$.
 To fit the experimental curves we use the two fitting parameters
that enter LAMH-theory, namely the critical temperature $T_c$ and
the transition width $\epsilon_c$. The agreement between theory
and experiment appears to be good, and our fitting procedure is
sensitive to $T_c$ and $\epsilon_c$. Indeed, when we plot a curve
where $\epsilon_c$ and $T_c$ deviate  by less than $10 \%$ from
the best fitting parameters ($A$ in Fig.~\ref{fg:et}) it is far
from the experimental curve $s1$.

 At low $T$ the experimental results deviate from the curve
calculated from the thermal-bosonic-LAMH-theory combined with the
quantum fermions $T_c$-reduction ($B$ in Fig.~\ref{fg:et}). For
$s1$ the deviation occurs below $R \approx 10 \Omega$. This could
be related to quantum phase slips.~\cite{SCW:Lau01} We note that
even for a perfect superconducting wire one should include the
contact resistance in "two-probe
measurements".~\cite{SCW:Tinkham96a} A rough estimate of the
contact resistance for sample $s1$  in Ref.~\cite{SCW:Bezryadin00}
shows that $R_{\rm contact} \approx R_q/(A/\lambda_F^2) \approx 11
\Omega$ (we take  $A \approx 50 nm^2$ and $ \lambda_F \approx 0.3
nm$). In Fig.~\ref{fg:et} the tail of the theoretical curve that
fits $s1$ splits into two (near the point mark by $B$). One curve
corresponds to LAMH theory and in the second we add a resistance
$R_0$ in series. We choose $R_0=6 \Omega$ to best fit the
experiment. Since the phase slip core action is exponentially
small in $A$ and $R_{\rm contact} \propto 1/A$ we expect the
former to be more relevant to $R_0$ for relatively narrow wires.

In addition, we show a curve that corresponds to a narrow long
wire, so that $R_N$ is similar to those in $s1$ and $s3$ (curves
$C$ and $D$) . Since the cross section $A$ is smaller both the
$T_c$-reduction is stronger and the contact resistance is higher.
In particular note that, as Ref.~\cite{SCW:Lau01} suggests, an
addition of $R_0$ in series to LAMH expressions does not predict a
negative curvature for $\log R(T)$ for all values of $T$. Due to
the $T_c$-reduction $R(T)$ at $T \sim 1.5K$ is very sensitive to
wire width.

Fig.~\ref{fg:fb} indicates that fermions determine $T_c$, and
thermal activation of bosonic phase slips determines the
resistance through the LAMH theory. The quantum bosons physics
should be relevant at low $T$. Thus, in contrast to the previous
theoretical~\cite{SCW:Zaikin97} calculations and
experimental~\cite{SCW:Bezryadin00} results, we do not expect
$R_N$ to be the only parameter that determines the quantum
transition.

While preparing this manuscript we  have learned of a wider set of
experiments  on superconducting wires of different lengths and
cross sections.~\cite{SCW:Lau01} They show, indeed, as was also
observed in Ref.~\cite{SCW:Kociak01}, that long wires become
superconducting even when their resistance is much higher than the
quantum resistance.~\cite{SCW:Lau01} A  bosonic
theory~\cite{SCW:Zaikin97} explains some features of this
behavior.
 We believe, however, that to fully understand the nature
of the phase transition in superconducting wires one should
consider the combined effects of fermionic and bosonic mechanisms.

\section{Conclusions.}

 We studied the interplay between fermions and bosons in
 superconducting wires. We argued that the disorder enhanced
 repulsion between the electrons -- the quantum physics of
 fermions -- reduces $T_c$ in dirty wires, with larger reduction for narrower wires.
  This quantum physics of
 fermions controls the parameters of the effective bosonic physics
 (both classical and quantum). A modification in wire width
 affect the low temperature behavior directly through the core
 energy of quantum phase slips and the contact resistance, and
 indirectly through the extra reduction of $T_c$ via the
 fermions mechanism. We showed that when analyzing the available
 experimental data it is important to include the reduction of
 $T_c$ together with quantum phase slips. The contact resistance in
 two-probe-measurements may not be neglected in some cases.

\section*{Acknowledgements}
It is a pleasure thank Daniel S.~Fisher and B.~I.~Halperin,
A.~Stern and R.~A.~Smith for valuable suggestions and C.~N.~Lau,
N.~Markovic, M.~Bockrath and A.~Bezeryadin and M.~Tinkham for
enlightening conversations and for allowing us to plot parts of
their data in this manuscript.


\begin{thebibliography}{99}
\bibitem{Remark2001}
The theory does not address the knee in the experimental data at
$~5.5K$ which is related to the superconducting phase transition
in the two dimensional leads.
\end{thebibliography}
\end{document}